\documentclass[prl,aps,prl,amssym,nofootinbib,floatfix,notitlepage]{revtex4-1} 
\usepackage{epsfig,natbib}
\usepackage{graphicx}
\newcommand{\be}{\begin{equation}}
\newcommand{\ee}{\end{equation}}
\newcommand{\bea}{\begin{eqnarray}}
\newcommand{\eea}{\end{eqnarray}}

\begin{document}
\title{Towards a Rigorous Proof of Magnetism  on the Edges of Graphene Nano-ribbons}
\author{Hamed Karimi and Ian Affleck}
 \affiliation{Department of Physics and Astronomy, University of British 
Columbia, Vancouver, B.C., Canada, V6T 1Z1}
\date{\today}

\begin{abstract}
A zigzag edge of a graphene nanoribbon supports localized zero modes, ignoring interactions. Based mainly 
on mean field arguments and numerical approaches, it has been suggested that interactions can produce a large magnetic moment 
on the edges.  By considering the Hubbard 
model in the weak coupling limit, $U\ll t$, for bearded as well as zigzag edges, we argue for such a magnetic state, based on Lieb's 
theorem.  Projecting the Hubbard interactions onto the flat edge band, we then prove that resulting 1 dimensional 
model has a fully polarized ferromagnetic ground state.  We also study excitons and the effects of second neighbor hopping as well as a potential 
energy term acting on the edge only, proposing a simple and possibly exact phase diagram with the magnetic moment varying 
smoothly to zero. Finally, we consider corrections of second order in $U$, arising from integrating out the gapless bulk Dirac excitations. 
\end{abstract}
\maketitle

One of the many fascinating properties predicted for graphene\cite{Castro} is that a non-interacting nanoribbon with zigzag edges has bands of states with 
energy exponentially small in the ribbon width, localized at the edges \cite{Fujita}.
 Unzipping of carbon nanotubes has recently provided a technique for producing 
nano-ribbons with clean edges and Scanning Tunnelling Microscopy (STM) on such ribbons \cite{Tao} has shown evidence for interacting edge states. 
It has been proposed,  on the basis of mean field theory \cite{Fujita,Jung1,Jung2}, density functional theory \cite{Son1,Son2} and various numerical 
techniques \cite{Hikihara,Dutta,Feldner}, that Hubbard interactions 
may induce ferromagnetic order of the electrons in these edge states, with the moments on opposite edges ordering antiferromagnetically. 
Graphene edge magnetism looks promising for applications in nano-electronics \cite{Kim}. 
However, there is no experimental evidence for edge magnetism in graphene ribbons and it is a matter of debate whether it is sufficiently robust 
to occur in realistic models \cite{Kuntsmann}, thus motivating a deeper and more general understanding of its origins. 
One might try to regard this magnetism as an essentially one dimensional (1D) phenomenon, since it arises from edge states, 
but the 1D Hubbard model is known to have a non-ferromagnetic ground state at all doping, being antiferromagnetic at half-filling. 
Recently \cite{Schmidt,Luitz}, the effective 1D model obtained by projecting the Hubbard interactions onto the edge states was studied numerically, 
and argued to lead to ferromagnetic order of an isolated edge. 
An interesting limit in which to try to prove edge magnetism is the weak interaction limit of the Hubbard model, $U\ll t$, at half-filling. 
We take two steps towards proving edge magnetism in this limit. The first involves 
applying Lieb's theorem \cite{Lieb} to the contrasting cases of a nanoribbon with two zigzag edges (ZZ) versus a ribbon with 
one zigzag and one bearded edge (ZB). (See Figure 1.)  
Then we prove that the  projected 1D Hamiltonian has a fully polarized ferromagnetic ground state.  We also study numerically other properties of the 1D model, 
obtaining the electron or hole addition energy and showing 
that there are bound spin-1 excitons. We then consider two important particle-hole symmetry 
breaking perturbations: second neighbor hopping, $t_2$ (in the entire ribbon) and a potential energy, $V_e$, acting on the edge atoms only, 
with the chemical potential maintained at the Dirac points of the bulk dispersion relation. 
 Both perturbations lead \cite{Sasaki} to the same new term in the 1D Hamiltonian, $\propto t_2-V_e\equiv \Delta$. We argue that the fully polarized ground state 
survives up to a critical value of $|\Delta |$ of $O(U)$, beyond which the ground state may still be found exactly and has a smoothly 
decreasing edge magnetic moment. Integrating out the bulk excitations of the ribbon leads to both inter-edge and intra-edge  interactions. 
The inter-edge interactions $\propto U^2/(tW^2)$, where $W$ is the ribbon width, produce antiferromagnetic order for the ZZ ribbon 
but ferromagnetic order for the ZB ribbon. Intra-edge interactions, $\propto U^2/t$, exhibit only a mild  logarithmic singularity at low energies, arising 
from the gapless nature of the bulk Dirac excitations.

 \begin{figure}
\centering \includegraphics*[width=0.85\linewidth]{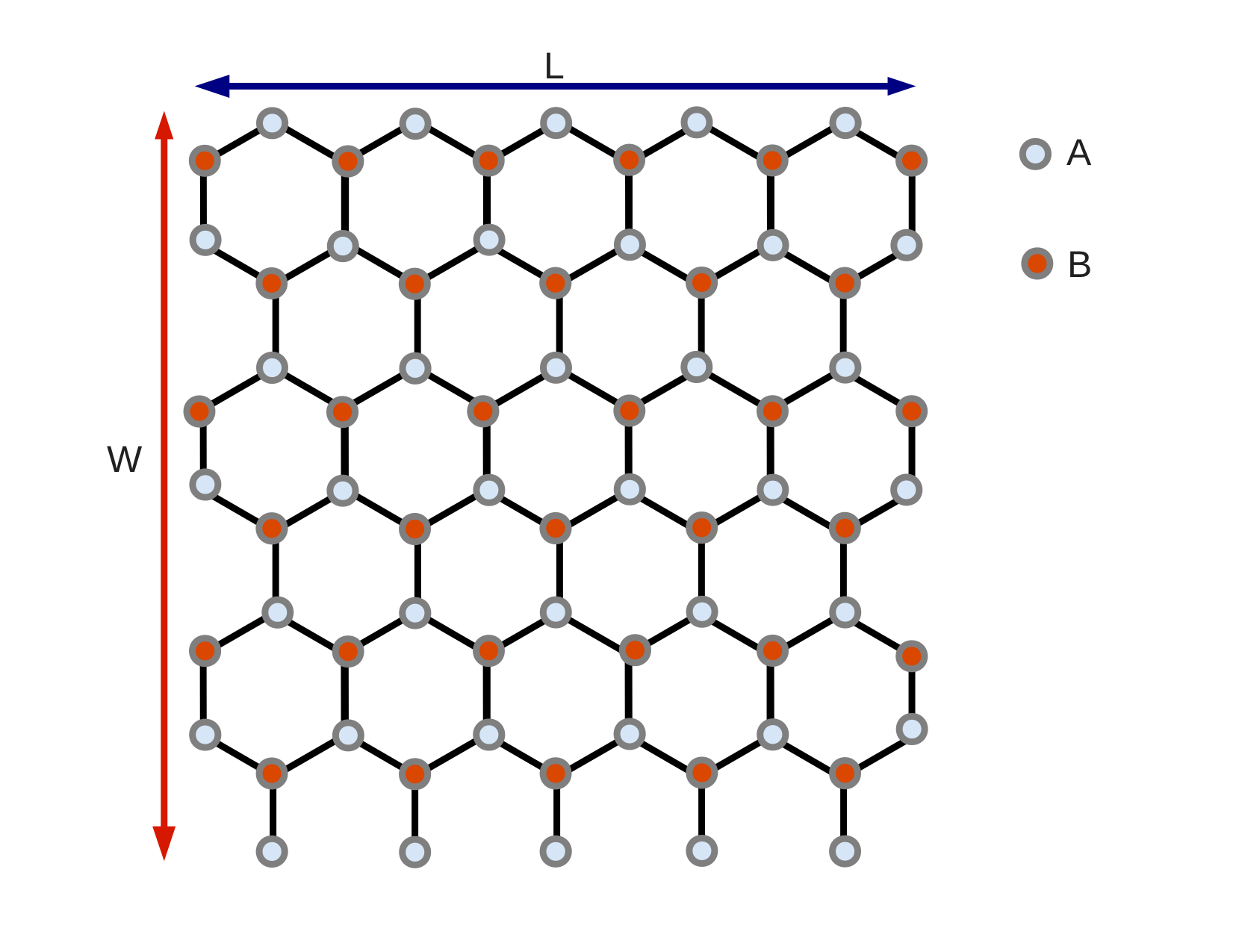}
\caption{A nanoribbon with an upper zigzag edge and lower bearded edge. In this example, $L=5$ and $W=6$. (We employ 
periodic boundary conditions in the $x$-direction.) }
\label{fig:ZB}
\end{figure}

We consider the Hubbard model at half-filling on a  long ribbon of honeycomb lattice
with periodic 
boundary conditions in the $x$-direction and zigzag edges.  We actually find it
convenient to first consider 
an upper zigzag edge and a lower bearded edge. [See Figure \ref{fig:ZB}.]  Let the number of
atoms along the 
zigzag edge, of $A$ type, be $L$. (Therefore the 
length of the ribbon is $\sqrt{3}aL$ where $a$ is the nearest neighbour separation. We generally set $\sqrt{3}a=1$.)  $L$ is
also the number of hairs in the beard, ending at $A$ sites. Noting that the number of $A$ sites 
minus the number of $B$ sites is $L$, it follows from Lieb's theorem \cite{Lieb}, that the
ground states have spin $S=L/2$ for all values of $U/t>0$.  We label the width of 
the strip by another positive integer, $W$ so that the width is $3aW/2$ in the ZB case 
or $a(3W+1)/2$ in the ZZ case. 
For the non-interacting model, $U=0$, and $W, L\gg 1$, there are approximately $L/3$
zero energy states localized at the upper zigzag edge 
and $2L/3$ localized at the lower bearded edge. The zigzag edge states
have wave-vectors, $2\pi /3\leq |k|<\pi$ while the bearded edge states
fill the rest of the Brillouin zone, $|k|\leq 2\pi /3$. It follows from
particle-hole symmetry that 
all of these states have exactly zero energy (for any $W$). These $L$ zero modes are
half-filled so we have a large degeneracy 
of ground states including a spin multiplet with spin $L/2$ obtained by filling each
of the $L$ states with a single electron with 
the same polarization. 
Now consider turning on a very small positive $U$, in the case of large $W$ and $L$. 
In the absence of boundaries we don't expect $U$ to have a large effect on the
ground state. A gapless non-magnetic Dirac liquid state is expected 
to persist up to a critical $U_c$ of $O(t)$. This follows from the fact that
perturbation theory in $U$ is infrared finite in the Dirac model; 
i.e. the 4-fermi interaction is irrelevant. Therefore, we expect the magnetic moment
to live on the edges.  The only physical  
explanation of the Lieb's theorem result seems to be that the fully polarized edge
state multiplet persists as the (unique) ground state as we turn on $U$.
This 
picture can be further substantiated by calculating the weak interaction between the
upper and lower edges, of order 
$U^2/(tW^2)$. This interaction is found to be ferromagnetic, as we show below.  Thus there must be a
spin $\approx L/6$ on the zigzag edge 
and $\approx L/3$ on the bearded edge, with these two spins coupled ferromagnetically. Now
consider replacing the bearded 
edge by a second zigzag edge.  Lieb's theorem \cite{Lieb} now implies a zero spin
ground state since 
we have equal numbers of $A$ and $B$ sites. Now, at $U=0$, we have approximately $L/3$ edge states on both
lower and upper edges. These mix 
to form two bands, with $2\pi /3\leq k\leq 4\pi /3$, with energies exponentially small in $W$
and symmetric around $E=0$. 
Ignoring inter-edge interactions, we expect spin $\approx L/6$ on both upper and lower edges. 
In this case, the intra-edge interaction of order $U^2/(tW^2)$  is
antiferromagnetic, implying a zero 
spin ground state consistent with the result from Lieb's theorem. 
  A further 
consistency check can be obtained by going smoothly between zigzag and bearded lower
edges by turning on the 
hopping term, $t'$ on the hairs. Lieb's theorem implies spin $L/2$ for all $t$, $t'$
and $U>0$. 
For $t'=0$ we have a ZZ ribbon together with $L$ decoupled sites sitting below
the lower edge. 
The ZZ ribbon has spin $0$ but we can obtain a state with total spin $L/2$ by
polarizing the electron spins at the decoupled sites. Although the zigzag ribbon has total spin $0$, for
large $W$ we expect that the upper and lower 
edges have spin $\approx L/6$ with antiferromagnetic inter-edge coupling. Turning on $t'$
produces an effectively antiferromagnetic 
coupling between the spin $\approx L/6$ on the lower zigzag edge and $L/2$ on the nearly
decoupled sites. This  
gives a moment $\approx L/3$ which is now ferromagnetically coupled to the upper edge,
giving a total spin of $L/2$ 
as required by Lieb's theorem.

Assuming that the ground
state remains an unpolarized Dirac liquid up to $U=U_c$, the magnetism of the edges
seems to follow from Lieb's theorem for large $W$.  If the 
transition at $U_c$ is into a bulk antiferromagnetic state (with, for example, spin
up on $A$ sites and spin down on $B$ sites) 
then the edge magnetism should persist, since it is of this type, and may be
regarded as a 
sort of precursor of the bulk antiferromagnetic order.  The simplest possibility is that the
system goes from Dirac liquid into Mott-Hubbard insulator at $U_c$ 
but numerical evidence \cite{Meng} has been presented for a spin liquid phase at intermediate
$U$, of unknown edge magnetic properties. 

We note that the above arguments also apply   to carbon nanotubes \cite{Schmidt2}. Indeed,
since we have been considering 
periodic boundary conditions in the $x$-direction, we have actually been discussing
tubes, of circumference $L$ and length $W$. 
The magnetic moments exist on the upper and lower caps (i.e. rings) of the nanotubes
with ferromagnetic or antiferromagnetic 
inter-ring coupling for a bearded or zigzag lower ring respectively. (The half-filled 
bulk of the nanotube might be in a 1D version of a Mott-Hubbard insulating state 
but this only serves to weaken the effects of bulk states on edge states.)

We may further substantiate this picture by considering \cite{Schmidt} the weak intra-edge
interactions of $O(U)$. 
Simply projecting the Hubbard interaction onto the zero energy states on the zigzag
edge, in the 
large $W$ limit, gives a
Hamiltonian:
\bea
\mathcal{H}  = {1 \over 2}\sum_{k,k',q}\Gamma(k,k',q)\left[\sum_{\sigma}e^{\dagger}_{\sigma}(k+q)e_{\sigma}(k) -\delta_{q,0}\right]\left[\sum_{\sigma'}e^{\dagger}_{\sigma'}(k'-q)e_{\sigma'}(k') -\delta_{q,0}\right]+E_0\label{H1D}
\eea
Here $e_\alpha (k)$ annihilates an electron in an edge state with momentum $k$ and spin $\alpha = \uparrow$ or $\downarrow$. 
The interaction function is:
\be \Gamma (k,k',q) = {\left\{\left[ 1-\left(2\cos{k \over 2 }\right)^2\right]\left[ 1-\left(2\cos{k+q \over 2 }\right)^2\right]
\left[ 1-\left(2\cos{k' \over 2 }\right)^2\right]\left[ 1-\left(2\cos{k'-q \over 2 }\right)^2\right]\right\}^{1/2}
\over 1-16\cos{k \over 2 }\cos{k+q \over 2
}\cos{k'-q \over 2 }
\cos{k' \over 2 }  }.\label{Gamma}\ee
 The sum over $k$, $k'$ and $q$ is restricted to the 
band in which $2\pi /3<k$, $k'$, $k+q$, $k'-q<4\pi /3$ 
and $\Gamma (k,k',q)$ is strictly positive \cite{bandedge}. Periodic boundary conditions in the $x$-direction imply 
$k=2\pi n/L$ so the number of wave-vectors $N\approx L/3$. 
Note that we are considering the case of half-filling 
in the entire lattice and that the edge Hamiltonian is therefore invariant under the 
particle-hole symmetry transformation:
\be e_{\alpha}(k)\leftrightarrow e^\dagger_{\alpha}(k).\ee
This is highly unusual since normally a particle-hole symmetry transformation
relates a particle 
and hole at different wave-vectors. Here with an exactly flat band, the particle and
hole 
operators occur at the same wave-vectors. 
It is important to note that $\Gamma (k,k',q)$ arose from summing the wave-function of the edge states over 
sites at arbitrary distance $n$ from the zigzag edge and can be written: 
\be \Gamma(l,k,q) = \sum_{n=0}^\infty \;g_n(k) \; g_n(l)\; g_n(l+q)\; g_n(k-q) 
\ee
where 
\be g_n(k) \equiv \theta(1-|2\cos{k \over 2}|)\sqrt{1-\left(2\cos{k \over
2}\right)^2} \left(2\cos{k \over 2}\right)^n.
\ee
Thus, dropping the constant $E_0$,  we may write:
\be \mathcal{H}  = {1 \over 2}\sum_{n,q}O^{\dagger}_n(q)O_n(q)\ee
with
\be O_n(q)\equiv \sum_k g_n(k)g_n(k+q)\left[\sum_{\sigma}e^{\dagger}_{\sigma}(k+q)e_{\sigma}(k) -\delta_{q,0}\right]\ee
It follows that all eigenstates of $H$ are non-negative.  It is can be seen that fully spin-polarized state is a zero 
energy eigenstate and therefore a ground state.  To check this consider, for example, the representative 
fully polarized state where all electron spins are in the up direction. Then clearly $O_n(q)$ annihilates this 
state for all non-zero $q$ since the spin up terms in $O_n(q)$ try to produce a spin up electron in an 
occupied state with wave-vector $k+q$ while the spin down terms try to annihilate a spin down 
electron in a vacant state of wave-vector $k$. $O_n(0)$ also annihilates this state since 
the occupancy of each single particle state is precisely $1$. 

It is also possible, though more difficult, to argue that the  fully polarized multiplet, of spin $S=L/6$, are 
the unique groundstates of the projected 1D Hamiltonian. 
 To prove that fully polarized states are the unique ground states of $H$ we need to prove that the only 
 states annihilated by $O_n(q)^\dagger O_n(q)$ for all $n$ and $q$ are fully polarized (that is, have maximal total spin). 
For convenience, in this paragraph, we take all momenta to be in the region of $[-\pi/3,\pi/3]$, which can be obtained by shifting all of them by $\pi$. Suppose $ |\psi\rangle $ is such that for any $n,q$ , $O_n(q) |\psi\rangle = 0$. Then we have
\bea
& & \sum_k g_n(k)g_n(k+q)\left[e^{\dagger}_{\sigma}(k+q)e_{\sigma}(k) -\delta_{q,0}\right]|\psi\rangle = 0 \nonumber \\
& = & \sum_{k>0} g_n(k)g_n(k+q)\left[e^{\dagger}_{\sigma}(k+q)e_{\sigma}(k)+e^{\dagger}_{\sigma}(-k)e_{\sigma}(-k-q) -2\delta_{q,0}\right]|\psi\rangle = 0
\eea
since $g_n(k)g_n(k+q)=g_n(-k)g_n(-k-q)$.  (Repeated spin indices are summed in this section.)
For fixed $q$ and using the definition of $g_n$ we have, for any $n$ that
\be 
\label{eq:n} 
\sum_{k>0} \left(4 \sin({k \over 2})\sin({k+q \over 2})\right)^n |\psi^{(q)}_k\rangle = 0 \ee
\be |\psi^{(q)}_k\rangle \equiv \sqrt{1-(2\sin k/2)^2}\sqrt{1-(2\sin (k+q)/2)^2} \left[e^{\dagger}_{\sigma}(k+q)e_{\sigma}(k)+e^{\dagger}_{\sigma}(-k)e_{\sigma}(-k-q) -2\delta_{q,0}\right]|\psi\rangle\ee
Since $n$ runs from $0$ to $\infty$,   the number of independent momenta is $L/3$ and all the $\left(4 \sin({k \over 2})\sin({k+q \over 2})\right)$ are different,  the determinant of the Vandermonde matrix is non-zero so Eq. (\ref{eq:n}) is satisfied if and only if for any $k,q$ we have
\be 
\label{condition}
\left[e^{\dagger}_{\sigma}(k+q)e_{\sigma}(k)+e^{\dagger}_{\sigma}(-k)e_{\sigma}(-k-q) -2\delta_{q,0}\right]|\psi\rangle = 0.\label{psiq}
\ee 
First, using Eq. (\ref{psiq}) for $q=0$, we get $n(k)+n(-k) = 2$, thus the  only possible terms have $n(k)=n(-k)=1 $ or $n(k)=0 $ and  $n(-k)=2 $ or vice versa.
In general $|\psi\rangle$ could be written as a linear combination of Fock states $\prod c^{\dagger}_{\sigma}(k)|0>$. We first will show that in the expansion of $|\psi\rangle$ in terms of such states there is no Fock state which for any momentum $k$ we have a vacancy or double occupancy in that momentum state. In other words, in the expansion of $|\psi\rangle$, with condition (\ref{condition}) and $n(k)+n(-k) = 2$, only Fock states with  singly occupied momentum states are allowed.\\
Suppose that there is a state which has the property  $n(k)+n(-k) = 2$ for any $k$ and has double occupancy at momentum $l$ (vacancy at momentum $-l$); call this state  $\phi$:
\be 
|\phi\rangle = |\cdots,\underbrace{0}_{-l},\cdots,\underbrace{\downarrow\uparrow}_l,\cdots\rangle
\ee 
and we suppose $\langle \phi | \psi \rangle \neq 0$. We impose the condition (\ref{condition}) on $|\psi\rangle$ for $q=-2l$ and $k=l$. Thus we should have $e^{\dagger}_{\sigma}(-l)e_{\sigma}(l)|\psi\rangle = 0$. Let us first look at the action of $e^{\dagger}_{\sigma}(-l)e_{\sigma}(l)$ on $\phi$,
\bea
\label{action}
 & & e^{\dagger}_{\sigma}(-l)e_{\sigma}(l)|\phi\rangle = e^{\dagger}_{\sigma}(-l)e_{\sigma}(l)  |\cdots,\underbrace{0}_{-l},\cdots,\underbrace{\downarrow\uparrow}_l,\cdots\rangle   \nonumber \\
& = &|\cdots,\underbrace{\uparrow}_{-l},\cdots,\underbrace{\downarrow}_l,\cdots\rangle -|\cdots,\underbrace{\downarrow}_{-l},\cdots,\underbrace{\uparrow}_l,\cdots\rangle 
\eea 
Now, in order to satisfy the condition $e^{\dagger}_{\sigma}(-l)e_{\sigma}(l)|\psi\rangle = 0$, we should have some other Fock states in the expansion of $|\psi\rangle$ such that the action of $e^{\dagger}_{\sigma}(-l)e_{\sigma}(l)$ on them could cancel the terms created in the second line of Eq. (\ref{action}). $e^{\dagger}_{\sigma}(-l)e_{\sigma}(l)$ only acts on states with momentum $l,-l$ thus does not change the spin configurations of the other (singly occupied) states. There are only three possible Fock states which have the same configurations of the singly occupied states: 
\bea
|1\rangle & = &  |\cdots,\underbrace{\downarrow\uparrow}_{-l},\cdots,\underbrace{0}_l,\cdots\rangle  \hspace*{0.5 in} e^{\dagger}_{\sigma}(-l)e_{\sigma}(l) |1\rangle = 0 \nonumber \\
|2\rangle & = &  |\cdots,\underbrace{\uparrow}_{-l},\cdots,\underbrace{\downarrow}_l,\cdots\rangle \hspace*{0.5 in} e^{\dagger}_{\sigma}(-l)e_{\sigma}(l) |2\rangle = |\cdots,\underbrace{\downarrow\uparrow}_{-l},\cdots,\underbrace{0}_l,\cdots\rangle  \nonumber \\
|3\rangle & = &  |\cdots,\underbrace{\downarrow}_{-l},\cdots,\underbrace{\uparrow}_l,\cdots\rangle \hspace*{0.5 in} e^{\dagger}_{\sigma}(-l)e_{\sigma}(l) |3\rangle  =-|\cdots,\underbrace{\downarrow\uparrow}_{-l},\cdots,\underbrace{0}_l,\cdots\rangle\eea 
We see that none of these states are able to cancel the terms created on (\ref{action}). Then the assumption is wrong and we have to have $\langle \phi| \psi \rangle = 0$.
Having proven this, we show that in the expansion of $|\psi\rangle$ in terms of singly occupied Fock states, only symmetric combinations like the following are acceptable
\be 
|\cdots,\underbrace{\uparrow}_{k_1},\cdots,\underbrace{\downarrow}_{k_2},\cdots\rangle + |\cdots,\underbrace{\downarrow}_{k_1},\cdots,\underbrace{\uparrow}_{k_2},\cdots\rangle
\ee
Suppose that, the Fock expansion of $|\psi \rangle$ has a term like $|\cdots,\underbrace{\uparrow}_{k_1},\cdots,\underbrace{\downarrow}_{k_2},\cdots\rangle$. 
 Now by chosing $k=k_1$ and $q=k_2-k_1$ we should have $(e^{\dagger}_{\sigma}(k_2)e_{\sigma}(k_1)
 +e^{\dagger}_{\sigma}(-k_1)e_{\sigma}(-k_2)|\psi \rangle = 0$. We also have 
\bea
 e^{\dagger}_{\sigma}(k_2)e_{\sigma}(k_1)|\cdots,\underbrace{\uparrow}_{k_1},\cdots,\underbrace{\downarrow}_{k_2},\cdots\rangle & = & |\cdots,\underbrace{0}_{k_1},\cdots,\underbrace{\downarrow\uparrow}_{k_2},\cdots\rangle \nonumber \\
  e^{\dagger}_{\sigma}(k_2)e_{\sigma}(k_1)|\cdots,\underbrace{\downarrow}_{k_1},\cdots,\underbrace{\uparrow}_{k_2},\cdots\rangle & = & -|\cdots,\underbrace{0}_{k_1},\cdots,\underbrace{\downarrow\uparrow}_{k_2},\cdots\rangle 
\eea
Thus we see that the symmetric combinations leads to zero, while the anti-symmetric ones give us a non-zero result.  
Thus $|\psi \rangle$ must be fully symmetric under exchanging all spins and is therefore of maximal spin. 

\begin{figure}
\centering \includegraphics*[width=0.85\linewidth]{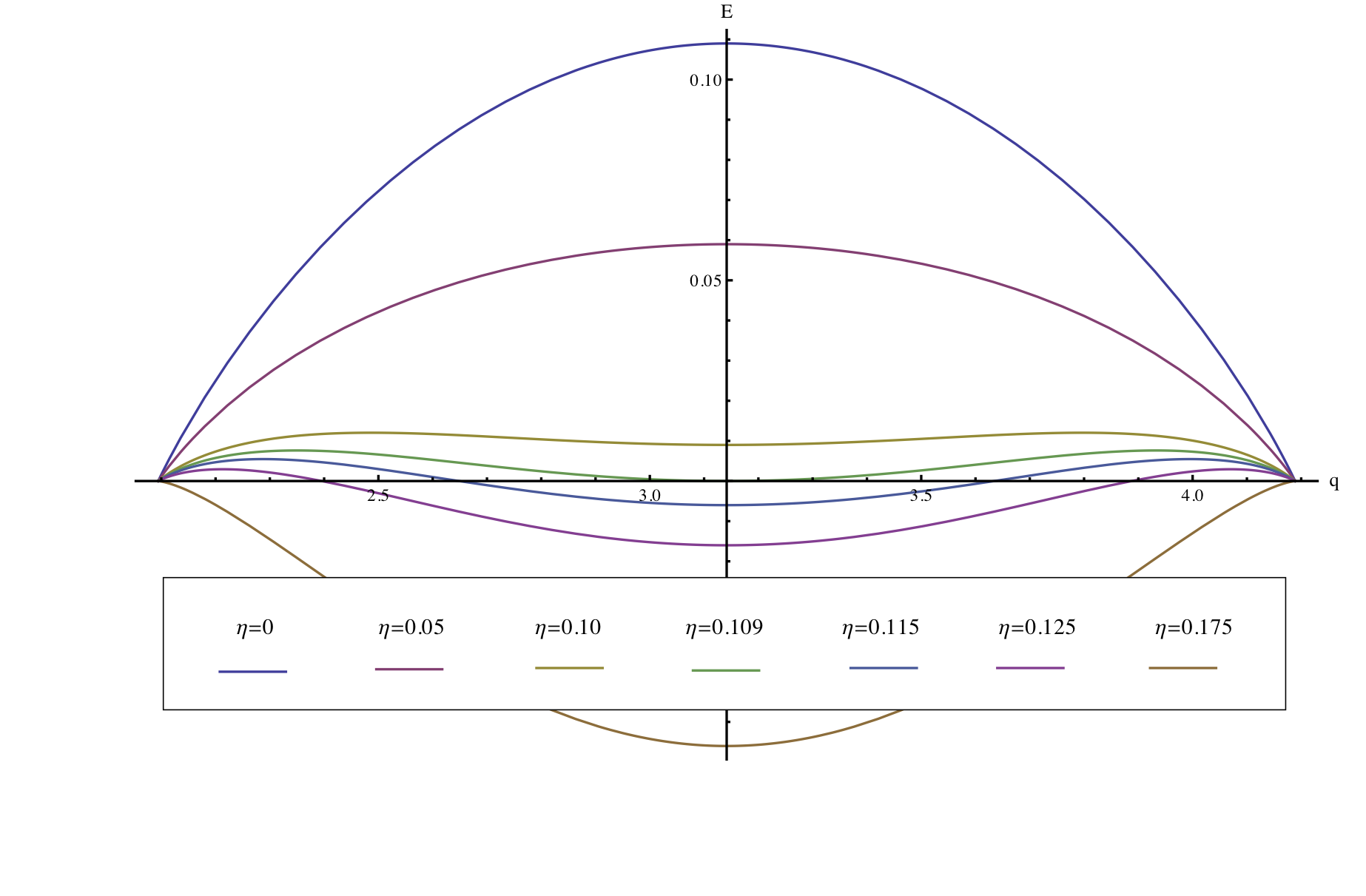}
\caption{Energy to add a spin down electron  of momentum $k$ for 
various values of $\eta \equiv \Delta /U \equiv (t_2-V_e)/U$.}
\label{fig:pe}
\end{figure}

One interesting quantity that follows from the Hamiltonian 
is the energy to add a spin down electron of momentum $k$, which is the same as the
energy to remove 
a spin up electron of momentum $k$:
\be \epsilon_k={U\over 2L}\sum_{k'}\Gamma (k,k',0).\label{ep}\ee
This quantity is plotted (at $L\to \infty$) in Fig. \ref{fig:pe}.  It vanishes linearly 
at the Dirac points, $k=2\pi /3$, $4\pi /3$. In principle, $\epsilon _k$ could be
measured in 
Angular Resolved Photo-emission Spectroscopy (ARPES) experiments. 
The corresponding electron addition or removal energy is given by $\epsilon_k$. 
The corresponding density of states $\propto 1/|d\epsilon_k /dk|$ could be 
measured by Scanning Tunnelling Microscopy (STM). 
With a spin-polarized STM tip and
an edge fully polarized 
in the $z$-direction, it would only be possible to tunnel in a spin-down electron or
tunnel out a spin-up electron. 

\begin{figure}
\centering \includegraphics*[width=0.85\linewidth]{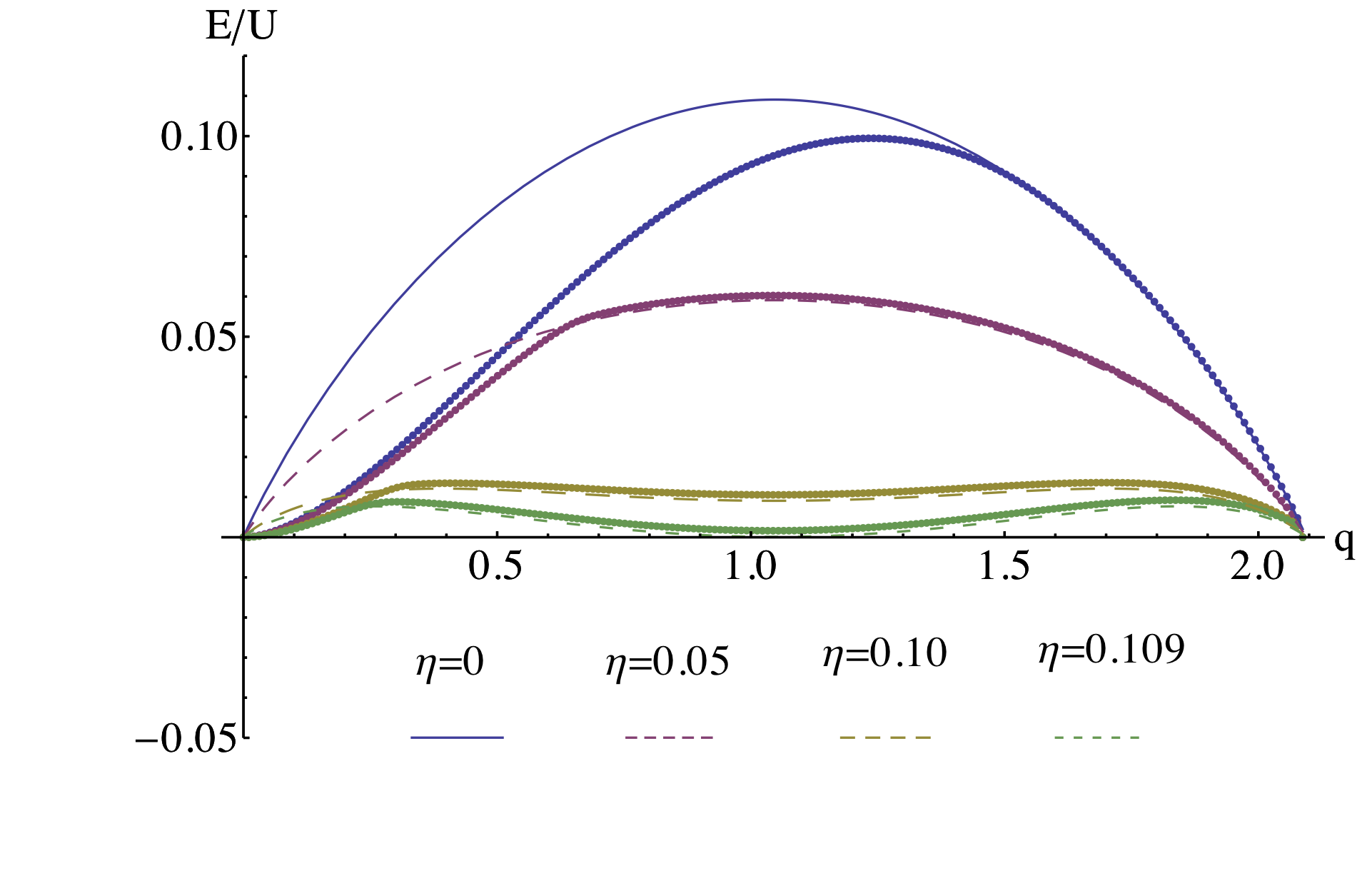}
\caption{Lowest energy particle-hole state (circles) and bottom of the particle-hole 
continuum (lines) for various values of $\eta \equiv \Delta /U\equiv (t_2-V_e)/U$. }
\label{fig:exciton}
\end{figure}

  We have calculated numerically the
lowest energy particle-hole 
 state of total momentum $q$, for $L$ up to $602$ ($N=200$). 
This is plotted in Fig. \ref{fig:exciton}
 along with the bottom of the particle-hole continuum. We see that a strongly bound
exciton exists for  most values of $q$, 
 as might be expected in this strongly interacting system. 
 However, the binding energy vanishing at $q=\pm 2\pi /3$. This vanishing can be
understood from the fact that $\Gamma (k,k',q)$ vanishes 
 when $k$ or $k'$ is at a band edge $2\pi /3$ or $4\pi /3$ so the zero energy particle and
hole become non-interacting at wave-vector 
 $2\pi /3$ and $-2\pi /3$ or vice versa.  
  
 While Lieb's theorem continues to imply a fully polarized ground state at
sufficiently small $U$ for any hopping 
 terms between opposite sub-lattices ($A$ to $B$), adding a small [O($U$)] second
neighbor hopping term, 
$t_2$, may destroy the fully polarized state.  Likewise, a single site potential,
$V_e$,  acting 
at the edge of the ribbon only, could destroy the fully polarized state. Temporarily
ignoring interactions, 
the zigzag edge states survive at finite $t_2$ and $V$ but develop a non-zero
dispersion given, 
to first order in $\Delta \equiv t_2-V_e$ , by \cite{Sasaki}:
$\epsilon_2(k)-\epsilon_F=\Delta (2\cos k +1)$. 
 breaking the particle-hole symmetry. Here we are assuming, for simplicity,
 that the bulk chemical potential is at the energy of the bulk Dirac points,  which
becomes $\epsilon_F=3t_2$. 
 (Shifting $\epsilon_F$ away from the Dirac points, the Hubbard interactions have a
larger effect in the bulk 
 rendering the edge model approach more questionable.) 
 Including 
a small $U$, the energy 
to add a spin down electron or remove a spin up electron at momentum $k$ now becomes 
\be E_{p/h}(k)=\epsilon_k\pm \Delta (2\cos k +1).\label{Eud}\ee
respectively, where $\epsilon_k$ is given in Eq. (\ref{ep}). $E_p(k)$ is plotted in Fig. \ref{fig:pe} for 
several values of $\Delta$.  We see that for
$|\Delta | <\Delta_c \approx 0.109U$,
the energy to add an electron or hole remains positive, so the edge states remain
undoped.  A local minimum at $k=\pi$ develops in $E_p(k)$ for $\Delta >.087U$, 
and $E_p(k)$ becomes negative in the vicinity of $k=\pi$ for $\Delta >\Delta_c$. 
 The lowest energy of a particle-hole state, and the bottom of the particle-hole
continuum for various values of $\Delta$ are shown in  Fig. {\ref{fig:exciton}.   We
see that
 the exciton becomes unbound except for wave-vectors near zero, as $|\Delta |$
increases.
For  $|\Delta |>\Delta_c$ the edge states become 
 doped, adsorbing electrons or holes from the bulk.  The simplest assumption for $|\Delta
|>\Delta_c$ is that a Fermi sea of 
  spin-down electrons or spin up holes forms near $k=\pi$, for $\Delta >\Delta_c$ or $\Delta
<-\Delta_c$ respectively. This assumption is reasonable since there appear to be no bound excitons 
for $\Delta >\Delta_c$. We also calculated  for $\Delta$ near $\Delta_c$ and  $L\leq 74$, the lowest energy state with
$M=N/2-2$,
finding no states below the 2-particle, 2-hole continuum, consistent with this assumption. 
These are  exact eigenstates 
 of the Hamiltonian of Eq. (\ref{H1D}), (\ref{Eud})  with the added holes
or particles non-interacting.  The corresponding 
 exact result for magnetization versus $\Delta$ is plotted in Fig. \ref{fig:M}, given this
assumption. 
 The non-interacting nature is a simple consequence of the fact that the on-site
Hubbard model only gives interaction between 
 electrons of opposite spin. On the other hand, we cannot rule out the possibility 
 that the ground state for $\Delta
>\Delta_c$  contains a finite density of 
 spin-up holes as well as the spin-down electrons (and similarly for $\Delta
<-\Delta_c$).  In that case, Hubbard interactions 
 have a non-trivial effect.  
  In any event, adding a nearest neighbour Coulomb repulsion term to the bulk Hamiltonian has no
effect on 
 the projected edge Hamiltonian, since such a term acts between $A$ and $B$ sites
whereas the zigzag edge states live entirely 
 on one sublattice. On the other hand, a second neighbour Coulomb repulsion, $U_2$,
produces interactions between electrons with 
 parallel spins in the projected edge Hamiltonian.  While this doesn't change our
conclusions qualitatively in the undoped phase, it will produce 
 interaction effects  in the ground state for the doped case even if contains only
particles or only holes.  However, we might expect $U_2\ll U$, in which case these
effects could be quite small. 
 Thus in general we expect a one or two component Luttinger liquid for $|\Delta
|>\Delta_c$. 
  On the other 
 hand the edge phase occurring for  $|\Delta |<\Delta_c$ is definitely not a
Luttinger liquid. Instead, it
  might be described as a fully spin-polarized semi-metal since all levels 
  are filled with spin-up electrons and there is a 
 non-zero electron and hole addition energy for all wave-vectors accept the
band-edges, $2\pi /3$ and $4\pi /3$.
 
  \begin{figure}
\centering \includegraphics*[width=0.85\linewidth]{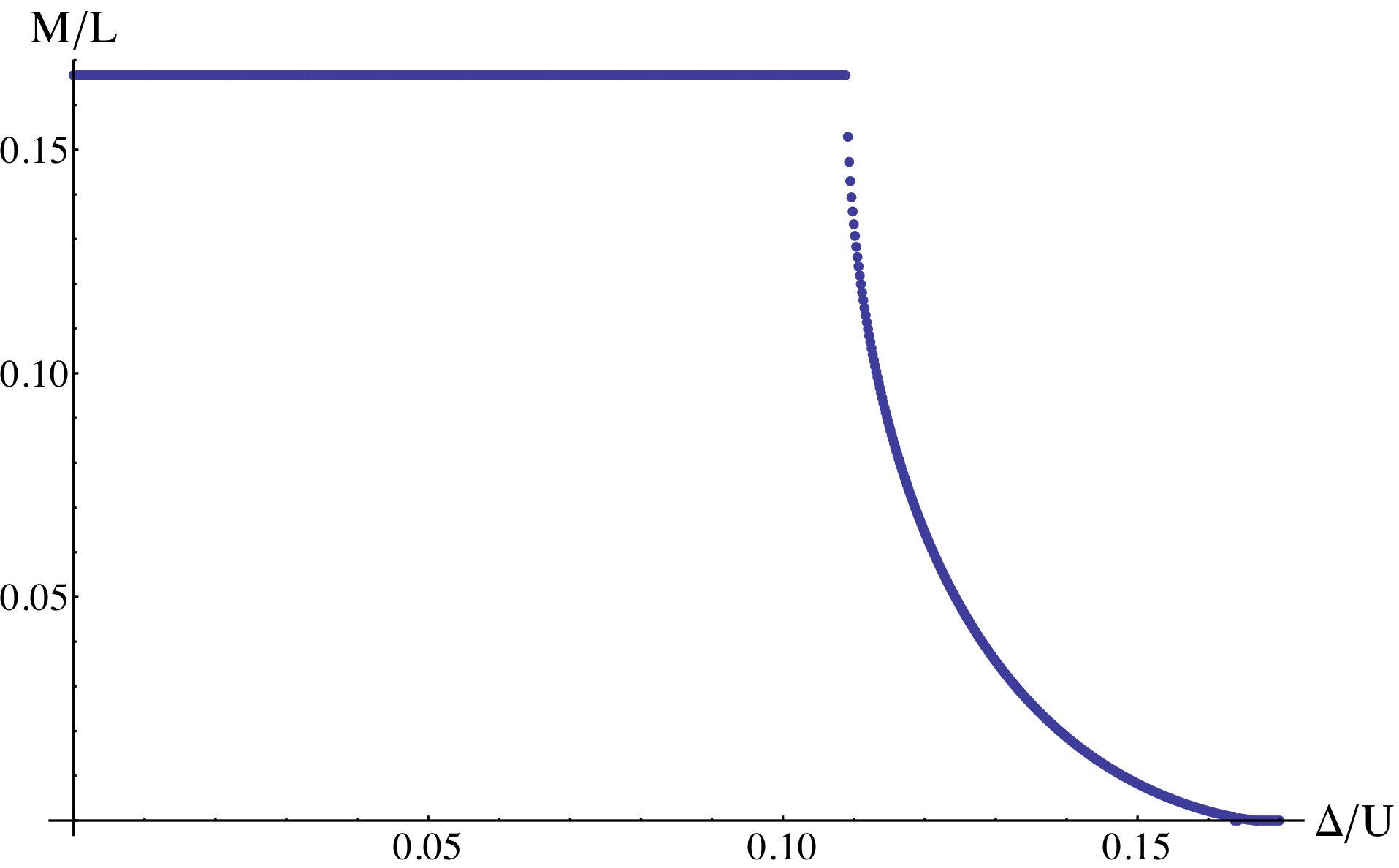}
\caption{Edge magnetization versus $\Delta \equiv t_2-V_e$. }
\label{fig:M}
\end{figure}

 There are also important effects of $O(U^2/t)$ which arise from the interactions
between bulk and edge states.  We can consider 
 integrating out the bulk states to obtain a low energy effective action for
the edge states. Due to the gapless nature 
 of the bulk Dirac spectrum, this produces long range retarded interactions among
the edge excitations. Decay processes of edge into bulk electrons are forbidden by energy-momentum conservation 
but the Feynman diagrams of Fig. \ref{fig:feynman} induce quartic interaction terms.
  For large $W$ and low energies we may calculate these interactions 
keeping only the low energy bulk states near the Dirac points, using the corresponding 
Dirac propagators. Note that we ignore interaction effects in the bulk, as discussed above. This is rather 
 similar to an RKKY interaction. The interaction 
 involving the dynamical spin operators \cite{other} $\vec S_{U/L}(\omega ,q)$ on
the upper and lower edge, respectively is:
 \be S_{\hbox{inter}}=\int {dqd\omega \over (2\pi )^2}\vec S_U(\omega
,q)\cdot\vec  S_L(-\omega ,-q)J_{\hbox{inter}}(\omega ,q,W)\ee
 where 
 \be J_{\hbox{inter}} (\omega ,q,W)=2\;U^2\int {d\omega 'dk\over (2\pi )^2}G(\omega ',k,0,W)G(\omega
-\omega ',q-k,0,W).\ee
 Here $G(\omega ,k,0,W)$ is the bulk free electron  Green's function with momentum
$k$ in the $x$-direction at $y=0$ and $y=W$ 
 with appropriate zigzag or bearded boundary conditions and projected onto the
sublattices corresponding to the 
 upper and lower edge ($A-A$ for zigzag-bearded or $A-B$ for zigzag-zigzag). 
 Using the linearized, Dirac dispersion relation, which is valid at small $1/W$, $\omega /t$ and $q$, 
 \bea G_{ZB}(\omega,k_x,y=0,y'=W)&=& {2iv^2_F\over W} \sum_n {(-1)^nk_n^2\omega\over \epsilon^2(k_x,k_n)
 [\omega^2+ \epsilon^2(k_x,k_n)]}\label{inter}\eea
 where $\epsilon (\vec k)=v_F|\vec k|$ is the Dirac dispersion relation.
 $G_{ZZ}$ is given by the same expression with $\omega$ replaced by $i\epsilon (k_x,k_n)$ in the numerator inside the sum. 
 The sum over $n$ can be taken up to an arbitrary ultra-violet cut-off whose value doesn't affect the 
 behavior at small $1/W$, $\omega /t$ and $q$. $k_n=\pi n/W$ for the ZB case. Although the 
 wave-vectors of edge modes are phase-shifted from these values in the ZZ case, this can be ignored 
 at leading order in $1/W$, allowing us to again use $k_n=\pi n/W$. 

 \begin{figure}
\centering \includegraphics*[width=0.85\linewidth]{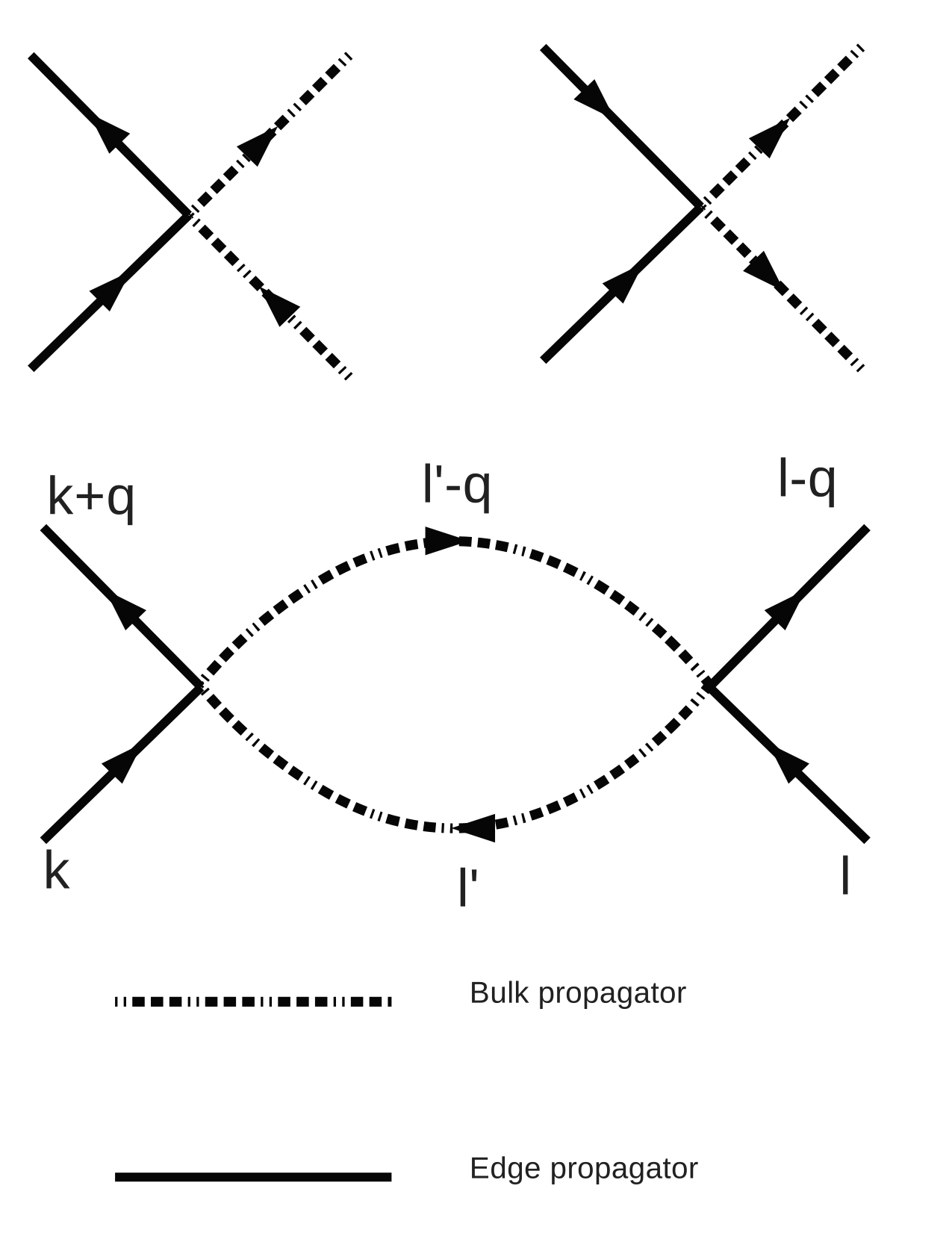}
\caption{Feynman diagrams inducing edge interactions from integrating out bulk states.}
\label{fig:feynman}
\end{figure}

It is straightforward to evaluate $J_{\hbox{inter}}(\omega ,q,W)$ numerically with
the two types of edges. The characteristic 
 scales for the $\omega$ and $q$ dependence of $J_{\hbox{inter}}$ are  set by $t/W$ and $1/W$, respectively.  Since the energy 
 scale of the inter-edge interaction in Eq. (\ref{inter}) is $U^2/(tW^2)$, it should
be permissible to ignore the retardation, 
 and evaluate $S_{\hbox{inter}}$ at $\omega =q=0$ to calculate the properties of low
energy states.  This gives:
 \bea J_{ZB/ZZ}(W)&=&\mp c {U^2\over t}{1\over W^2}\eea
 where the positive constant $c$ is given by the convergent   sums and integral:
 \be c\equiv {\sqrt{3} \over \pi }\times \sum_{n,m=1}^\infty (-1)^{n+m}\int_{-\infty}^\infty d\kappa {n^2m^2\over 
 (\kappa^2+m^2)(\kappa^2+n^2)}{1\over \sqrt{\kappa^2+m^2}+\sqrt{\kappa^2+n^2}}\approx 0.20\ee
 (A similar result was obtained in \cite{Jung1} for the ZZ case.)
We see that the ground state for the zigzag-bearded ribbon has spin $L/2$ while
that for the zigzag-zigzag case has spin 0, 
 as shown above rigorously using Lieb's theorem.  The remarkable fact that the
change in sign of this 
 tiny coupling drastically changes the spin of the ground state provides evidence
for the polarized nature 
 of the edge spins.  There is also a large manifold of low energy states, which 
 are simply the eigenstates of $J_{ZZ}\vec S_T\cdot \vec S_B$ with $S_T=S_B=L/6$ (in
the ZZ case). 
 
 Another important effect of $O(U^2/t)$ is the intra-edge interaction, independent of
$W$. For a zigzag edge by integrating out the low-energy bulk excitations, the spin part is:
\be S_{\hbox{intra}}=\int {dqd\omega \over (2\pi )^2}\vec S(\omega ,q)\cdot \vec 
S(\omega ,q)J_{\hbox{intra}}(\omega ,q)\ee
with
\be 
\label{jintra}
J_{\hbox{intra}}(\omega ,q)=2\;U^2\int_{k,\omega '/v_f < \Lambda} {d\omega 'dk\over (2\pi )^2}G(\omega ',k,y=y'=0)G(\omega -\omega
',q-k,y=y'=0)\ee
Now the free bulk Green's function, with zigzag edge boundary conditions, may be evaluated for a semi-infinite system, giving, at small $k_x$ (measured 
from a Dirac point) and small $\omega$:
\be G(\omega ,k_x,y=y'=0)\approx 2i v_F^2\int{dk_y \over 2\pi}{k_y^2\over(v_Fk)^2}{\omega
\over \omega^2+(v_Fk)^2}\ee
By using this green function, the $J_{\hbox{intra}}$ of  Eq. (\ref{jintra}) is ultraviolet divergence and the integral should be cut off at some point $\Lambda$. Although the 
resulting $J_{\hbox{intra}}(\omega,q)$ is cut off dependent, by ignoring the weak retardation, the corrections to the energy of the excitons is proportional to $ J_{\hbox{intra}}(0,q)-J_{\hbox{intra}}(0,0) $, which is cut off independent. For small $\omega$ and $q$ these $O(U^2)$ intra-edge interactions become more singular than the $O(U)$ terms 
by logarithmic factors of $q^2\ln q$, giving for example the final equation in the main paper. 

As mentioned above, Eq. (\ref{jintra}) only includes the effect of low energy bulk excitations; it is still possible that the high energy bulk excitations could wipe out this singularity of the exciton dispersion relation. By using the exact form of the bulk wavefunctions one can determine the exact  intra-edge interaction of $O(U^2)$. This has a more 
complicated form than Eq. (\ref{jintra}).  Nonetheless, it can be shown that the only part of this interaction which contributes to
 this $ \ln q$ singular correction to the excition dispersion relation is the contribution from low energy bulk excitations of the form of Eq. (\ref{jintra}).

We leave a more detailed study of these effects of $O(U^2)$ and higher for the future. A reasonable approach 
might be to ignore the bulk interactions, since they are irrelevant, but analyse the bulk-edge Hubbard interactions 
using the renormalization group. This  corresponds to a novel type of boundary critical phenomena in which 
the bulk is a massless (2+1) dimensional Dirac liquid and the edge is a one-dimensional spin-polarized 
semi-metal. The arguments based on Lieb's theorem imply that the edge magnetic moment remains stable 
against weak interactions. 

\begin{acknowledgments}We would like to thank Ion Garate for his collaboration in the early stages of this project 
and Pawel Hawryak for interesting discussions. 
The research was supported in part by NSERC and CIfAR. We also thank the Galileo Galilei Institute where 
this work was completed. 
\end{acknowledgments}


\begin{thebibliography}{50}
\bibitem{bandedge}  In the case where $L$ is a multiple of $3$ there are states exactly at the band edges $\pm 2\pi /3$. We choose not 
to include them in the edge Hamiltonian since they are delocalized over the entire ribbon.
\bibitem{other} There are additional inter-edge interactions involving the number
operators and pair creation operators 
on upper and lower edges which will be discussed in a later article. However these
have little effect at large $W$.  
The spin term dominates  because the edges are spin polarized.
\end{thebibliography}
\end{document}